\documentclass{elsart3}
\usepackage{graphics}
\usepackage{epsfig}
\usepackage{amssymb}
\usepackage{cite}
\usepackage{subfigure}

\begin{document}
  \newcommand{\greeksym}[1]{{\usefont{U}{psy}{m}{n}#1}}
  \newcommand{\umu}{\mbox{\greeksym{m}}}
  \newcommand{\udelta}{\mbox{\greeksym{d}}}
  \newcommand{\uDelta}{\mbox{\greeksym{D}}}
  \newcommand{\uOmega}{\mbox{\greeksym{W}}}
  \newcommand{\uPi}{\mbox{\greeksym{P}}}
  \newcommand{\ualpha}{\mbox{\greeksym{a}}}
  \begin{frontmatter}


\vspace*{-11mm}{\it \begin{flushleft} \small
Talk presented at the Workshop on Tracking in high Multiplicity Environments (TIME 2005),\\
October 3-7 2005, Z\"urich,Switzerland.
\end{flushleft}}
\title{Simulation and hit reconstruction \\ of irradiated pixel sensors for the CMS experiment}

\author[uniz]{E.~Alag\"oz},
\author[uniz]{V.~Chiochia\corauthref{cor1}}, \ead{vincenzo.chiochia@cern.ch}
\author[jhu]{M.~Swartz}

\corauth[cor1]{Corresponding author}
\address[uniz]{Physik Institut der Universit\"at Z\"urich-Irchel, 8057 Z\"urich, Switzerland}
\address[jhu]{Johns Hopkins University, Baltimore, MD 21218, USA}
\begin{abstract}
In this paper a detailed simulation of irradiated pixel sensors
was used to investigate the effects of radiation damage on the 
position determination and optimize the hit reconstruction algorithms.
The simulation implements a model of radiation damage by including two
defect levels with opposite charge states and trapping of charge carriers.
The simulation shows that a position resolution below 15 $\mu$m along the CMS $r-\phi$ plane can 
be achieved after an irradiation fluence of $5.9\times10^{14}$ n$_{\rm{eq}}/$cm$^2$.
In addition, we show that systematic errors in the position determination
can be largely reduced by applying $\eta$ corrections.
\end{abstract}

\end{frontmatter}

\section{Introduction}
The CMS experiment, currently under construction at the Large Hadron Collider
(LHC) will include a silicon pixel detector~\cite{CMSTrackerTDR:1998} to allow tracking in the region closest
to the interaction point. The detector will be a key component for reconstructing interaction
vertices and heavy quark decays in a particulary harsh environment, characterized by
a high track multiplicity and heavy irradiation.
At the full LHC luminosity the innermost layer, with a radius of 4.3 cm, 
will be exposed to a particle fluence of $3\times10^{14}$ n$_{\rm{eq}}/$cm$^2$/yr.
                                                      
In order to evaluate the effects of irradiation and optimize the algorithms
for the position determination a detailed  simulation of the pixel sensors
was implemented.
In~\cite{Chiochia:2004qh} we have proved 
that it is possible to adequately describe
the charge collection characteristics of a heavily irradiated silicon detector in terms
of a tuned double junction model which produces a doubly peaked electric field profile across the sensor.  
The modeling is supported by the evidence of doubly peaked electric fields obtained from beam
test measurements and presented in~\cite{Dorokhov:2004xk}.  
The dependence of the modeled trap concentrations upon fluence was presented in~\cite{Chiochia:2005ag}
and the temperature dependence of the model was discussed in~\cite{Swartz:2005vp}.
In this paper the simulation was used to study the position determination 
in irradiated pixel sensors.

This paper is organized as follows: the sensor simulation is described
in Section~\ref{sec:sensor_simulation}, in Section~\ref{sec:hit_rec} the
determination of the hit position in irradiated pixel sensors is discussed. 
The results are presented in Section~\ref{sec:results} and the conclusions
are given is Section~\ref{sec:conclusions}.

\section{Sensor simulation~\label{sec:sensor_simulation}}

The results presented in this paper rely upon a detailed 
sensor simulation that includes the modeling of irradiation effects in silicon.
The simulation, {\sc pixelav}~\cite{Swartz:2003ch,Swartz:CMSNote,Chiochia:2004qh}, 
incorporates the following elements: an accurate model of charge deposition by primary hadronic
tracks (in particular to model delta rays); a realistic 3-D intra-pixel electric field map; 
an established model of charge drift physics including mobilities, Hall Effect, and 3-D diffusion; 
a simulation of charge trapping and the signal induced from trapped charge; and a 
simulation of electronic noise, response, and threshold effects.  
The intra-pixel electric field map was generated using {\sc tcad 9.0} \cite{synopsys} to simultaneously 
solve Poisson's Equation, the carrier continuity equations, and various charge transport models.  

The simulated devices correspond to the baseline sensor design for the CMS
barrel pixel detector. The sensors are ``n-in-n'' devices, designed to collect
charge from n$^+$ structures implanted into n- bulk silicon.
The simulated samples were 22x32 arrays of 100x150 $\mu$m$^2$ pixels. 
The substrate was 285 $\mu$m thick, n-doped silicon. The donor concentration
was set to $1.2\times10^{12}$ cm$^{-3}$ corresponding to a depletion
voltage of about 75 V for an unirradiated device. 
The 4 T magnetic field was set as in the CMS configuration 
and the sensor temperature to -10$^\circ$ C.
The simulation did not include the ``punch-through'' structure on the n$^+$ implants which is used
to provide a high resistance connection to ground and to provide the possibility
of on-wafer IV measurements. 

The effect of irradiation was implemented in the {\sc tcad} simulation by including
two defect levels in the forbidden silicon bandgap with opposite
charge states and trapping of charge carriers. The activation energies
of the donor and acceptor traps were set to $(E_{V}+0.48)$ eV and $(E_{C}-0.525)$
eV, respectively, where $E_{V}$ and $E_{C}$ are the valence and conduction
band energy level, respectively. The trap densities and the capture cross sections 
for electrons and holes were obtained by fitting the model to beam test data 
as described in~\cite{Chiochia:2004qh,Chiochia:2005ag}. The simulated irradiation fluences
were $\Phi = 2 \times 10^{14}$ n$_{\rm{eq}}$/cm$^2$ and $\Phi = 5.9 \times 10^{14}$ n$_{\rm{eq}}$/cm$^2$
and the reverse bias was set to 200 V and 300 V, respectively.

\begin{figure}[hbt]\label{fig:e_field}
  \begin{center}
    \mbox{
      \subfigure[]{\scalebox{0.47}{
          \epsfig{file=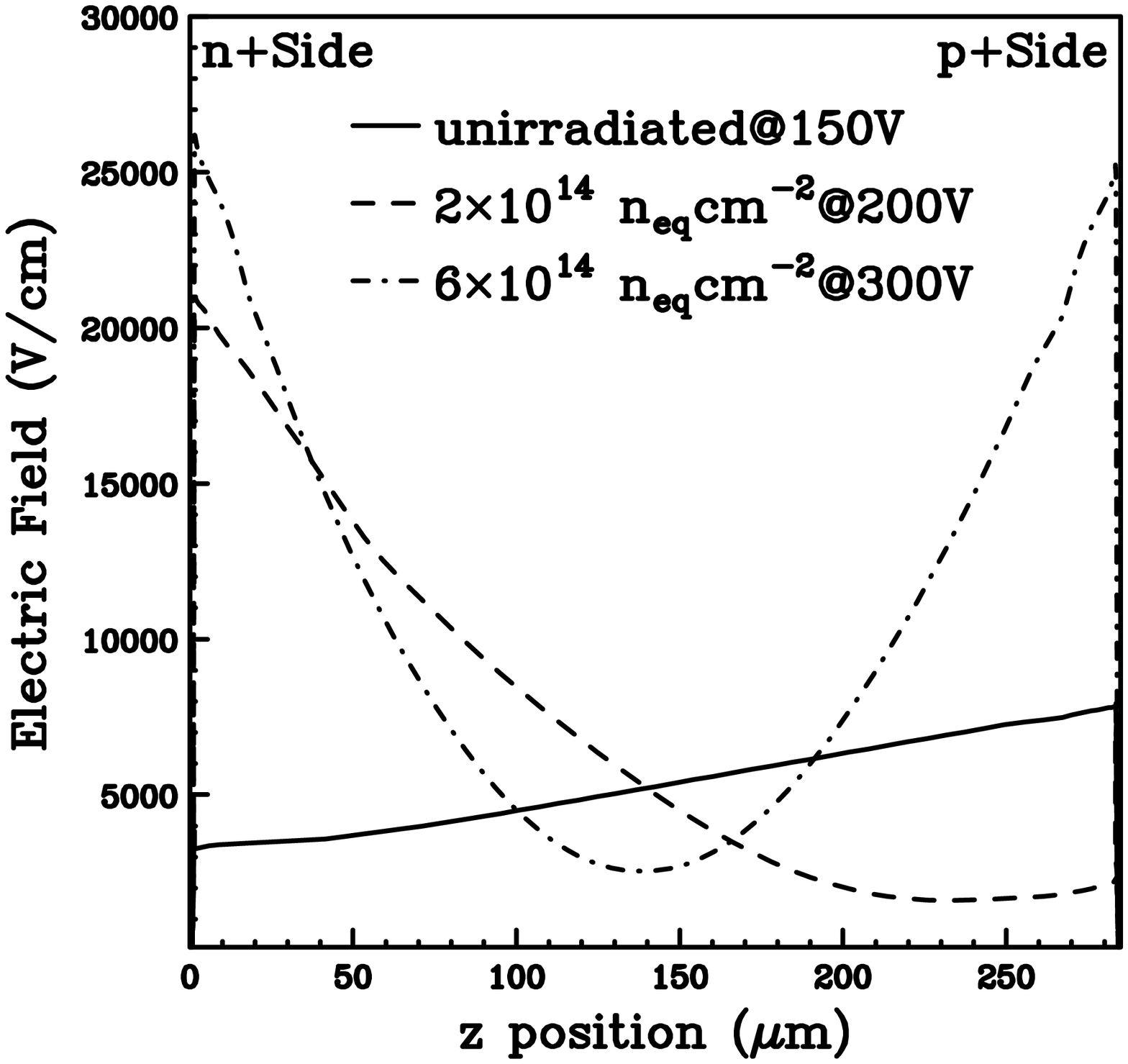,width=\linewidth}
          \label{fig:electric_field}
      }}
      \subfigure[]{\scalebox{0.50}{
          \epsfig{file=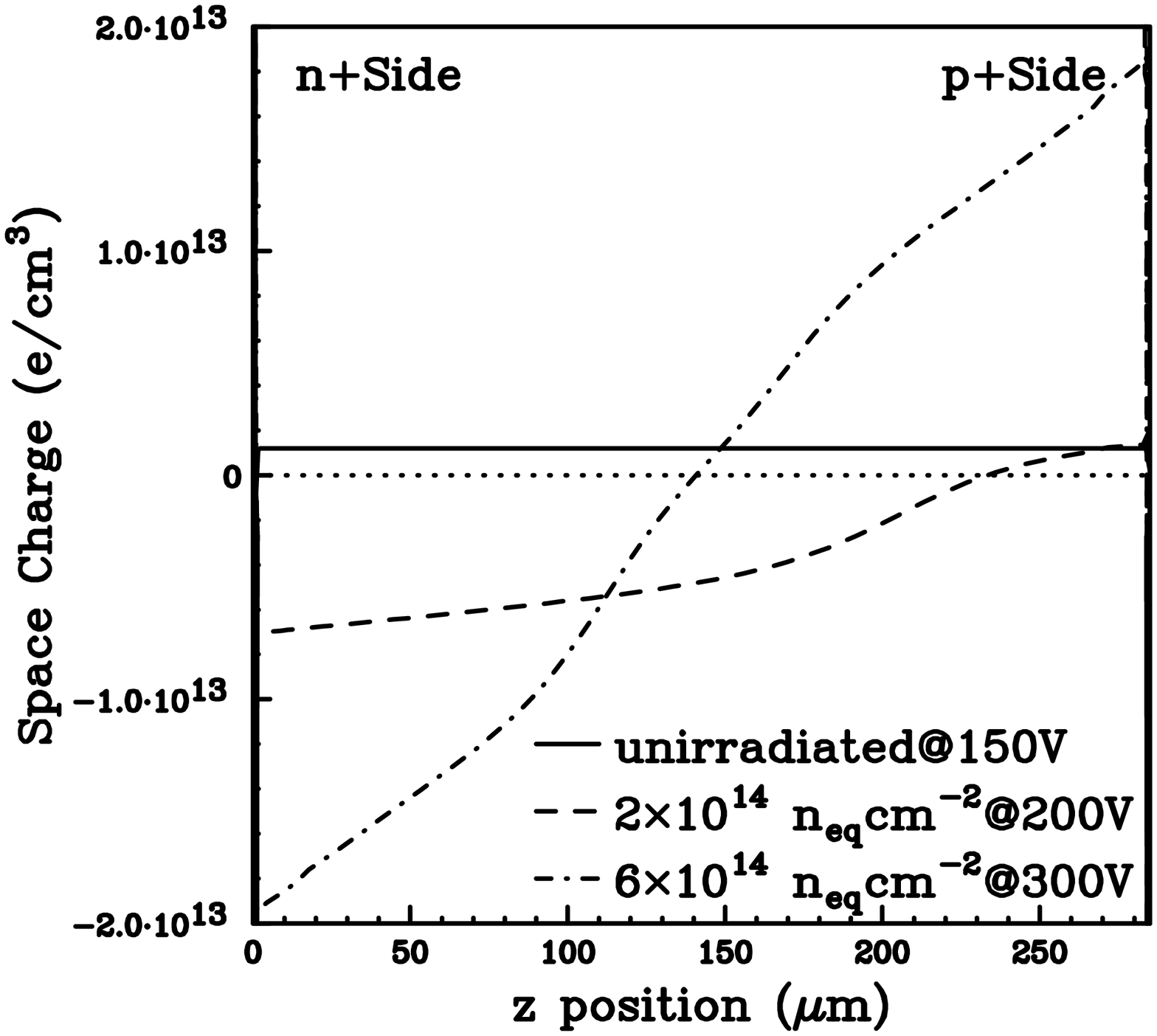,width=\linewidth}
          \label{fig:space_charge}
      }}
    }
    \caption{The $z$-component of the simulated electric field (a)
and space charge density (b), resulting from the two-trap model, are shown 
as a function of $z$ for unirradiated and irradiated devices.}
  \end{center}
\end{figure}

The electric field profile as function of the sensor depth is
shown in Fig.~\ref{fig:electric_field}.
The field has maxima at the detector implants and a minimum near the 
midplane which shifts towards the p$^+$ implant at lower 
fluences. The dependence of the space charge density upon the detector
depth is shown in Fig.~\ref{fig:space_charge}. Before irradiation the
sensor is characterized by a constant and positive space charge density
across the sensor bulk. At a fluence $\Phi = 2 \times 10^{14}$ n$_{\rm{eq}}$/cm$^2$
the device shows a negative space charge of about $-6 \times 10^{13}$ cm$^{-3}$
for most of its thickness, a compensated region corresponding
to the electric field minimum and a positive space charge density
close to the backplane. The space charge density and electric
field near the p$^+$ implant increase with the irradiation fluence.

\section{Hit reconstruction\label{sec:hit_rec}}

The spatial resolution of the pixel sensors is mainly determined by the readout pitch
and charge sharing between neighboring cells.
Pixels have a weak capacitive coupling and charge sharing is mainly due to diffusion
and drift of charge carriers under the combined effect of the magnetic and
electric fields. After irradiation, free carriers trapping produces an
inhomogeneous charge collection across the bulk and charge sharing between
neighboring pixels becomes strongly non linear on the impact position.
In addition, the beneficial effect of the Lorentz deflection is
reduced when a higher bias voltage is applied to provide a sufficient
drift field. In what follows we discuss measurements of the sensor
spatial resolution along the $r-\phi$ direction, where the
charge drift is affected by the Lorentz deflection.

\begin{figure}[hbt]
  \begin{center}
    \resizebox{0.9\linewidth}{!}{\includegraphics{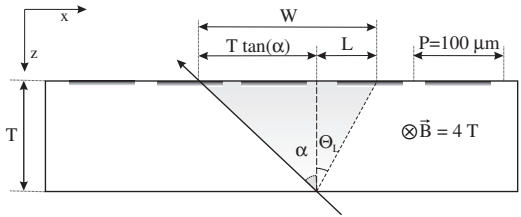}}
    \caption{Determination of the impact position in the transverse plane.}
    \label{fig:pos_resolution}
  \end{center}
\end{figure}

To reconstruct hits
pixels with charge above 2000 electrons were selected and clusters were formed
by adjacent pixels above threshold. Both side and corner adjacent pixels were
included in the cluster. In addition, clusters adjacent to the matrix border
were excluded.
Figure~\ref{fig:pos_resolution} shows the definition of the track impact angle $\alpha$
with respect to the sensor plane along the $x$ direction. The track is orthogonal
to the sensor plane along the axis orthogonal to $x$.
The magnetic field produces a Lorentz shift $L=T \tan(\Theta_L)$ towards the right direction.
Thus, the total charge width is given by
\begin{equation}
W = L - T \tan(\alpha).
\end{equation}
The cluster is projected along the $x$ axis by summing the charge collected
in the pixels with the same $x$ coordinate. If the cluster is given by a single pixel
its position is given by the pixel center. For larger clusters
the hit position is calculated with a center-of-gravity algorithm.

\begin{figure*}[hbt]
  \begin{center}
    \subfigure[]{ 
      \resizebox{0.40\linewidth}{!}{
	\includegraphics{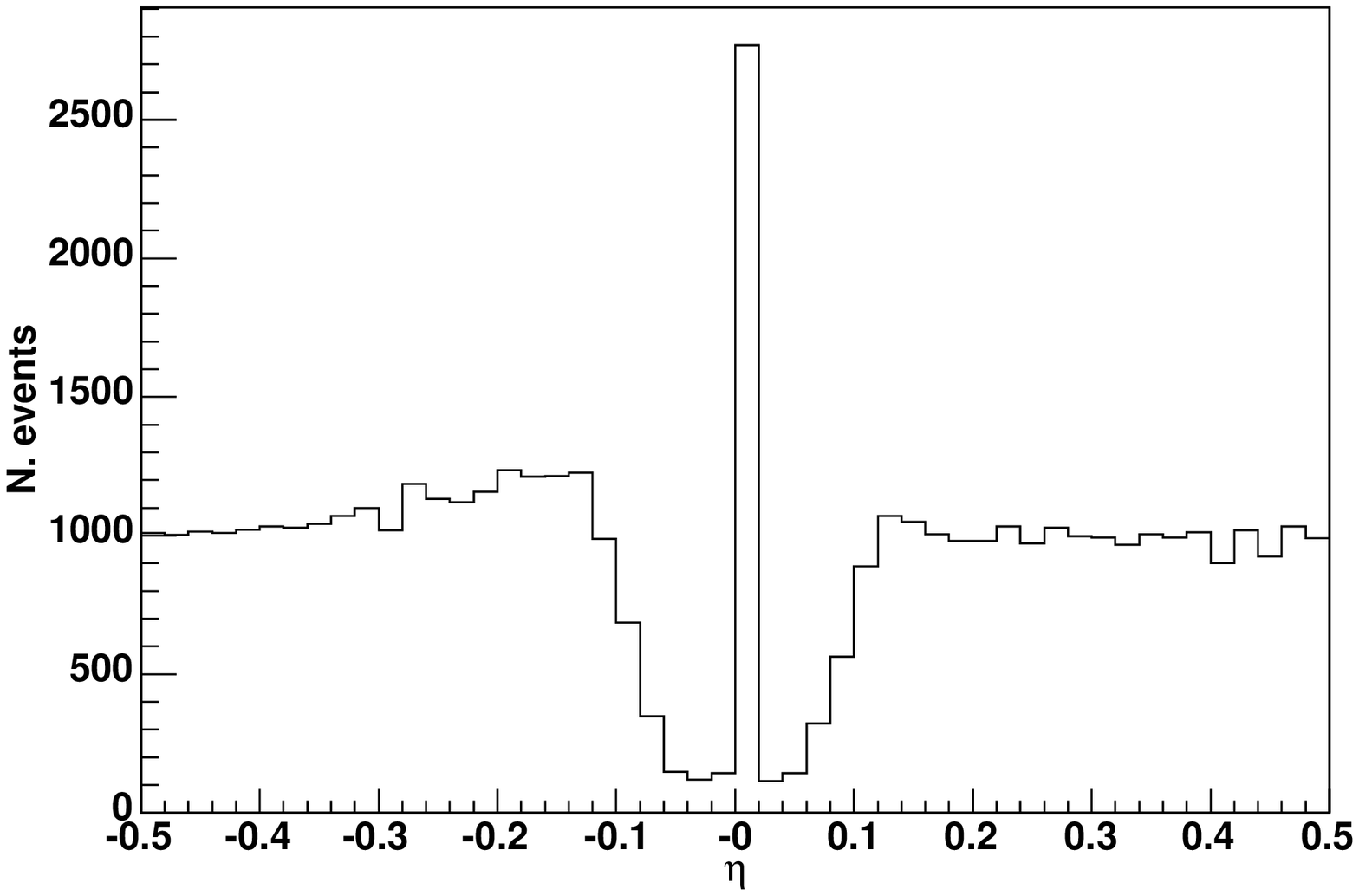}}
    }\vspace{-.4cm}
    \subfigure[]{ 
      \resizebox{0.40\linewidth}{!}{
	\includegraphics{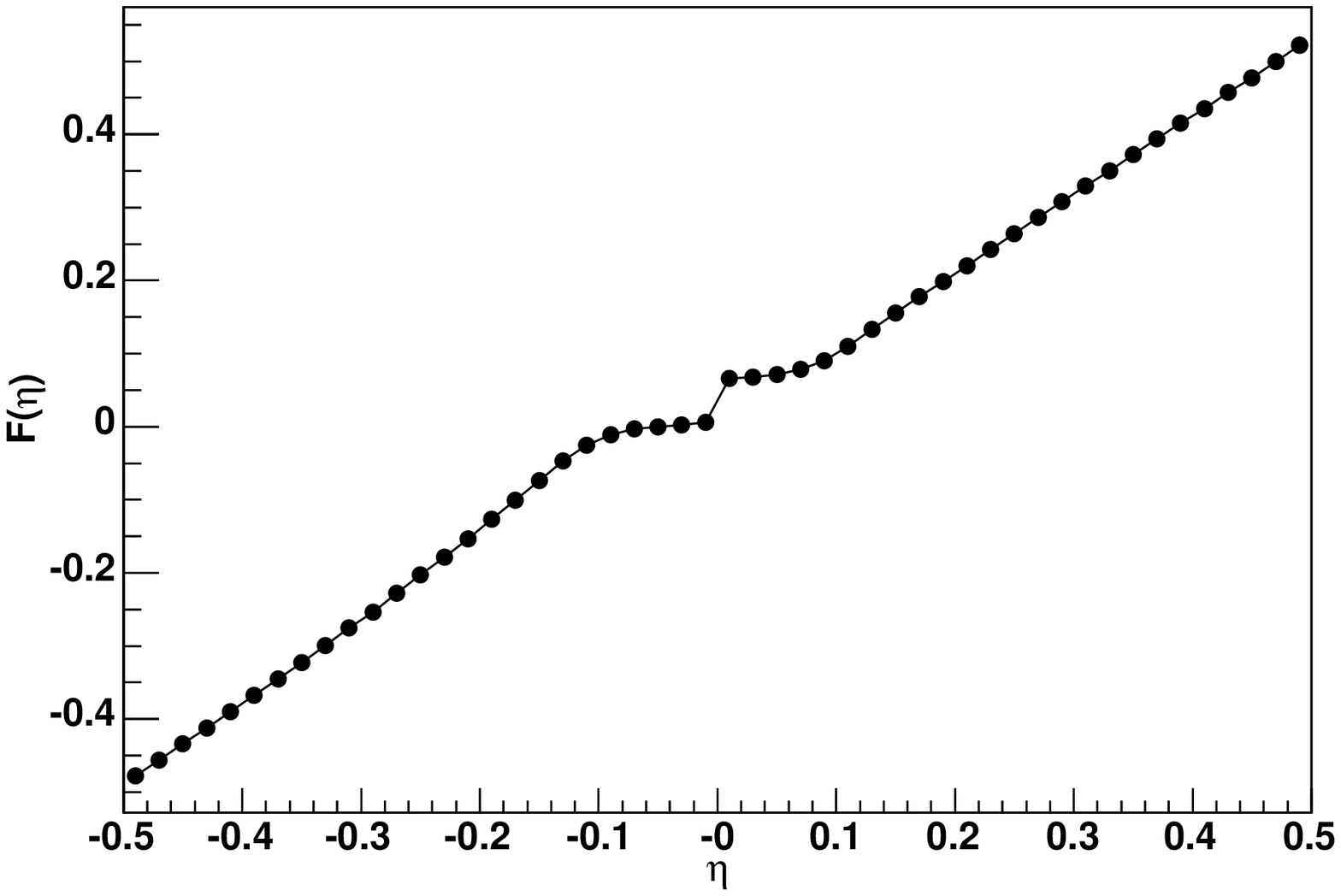}
      }
    }
    \caption{(a) Distribution of the reconstructed impact position within a single
pixel for perpendicular tracks. (b) Correction function $F(\eta)$ (see
Eq.~\ref{eq:F_eta}).}
    \label{fig:eta_function}
  \end{center}
\end{figure*}

To further improve the spatial resolution for events in which charge is shared
among several pixels the so-called $\eta$-correction is introduced~\cite{Belau:1983eh}.
As we will discuss, the correction is particularly effective on irradiated devices, where
the effects of inhomogeneous charge collection are larger.
Assuming that the number of particles crossing the sensor is uniformely distributed along $x$
we should expect the reconstructed position within a pixel to be uniformely distributed.
We define $\eta$ as the non-integer
part of the reconstructed position.
Figure~\ref{fig:eta_function} shows the distribution of $\eta$
for all events, where $\eta=0$ corresponds to the center of the pixel cell
and $\eta= \pm 0.5$ to the borders.
The measured distribution is almost flat in the pixel regions closer to the
pixel borders and it shows a dip at the center. The peak at $\eta=0$ is due
to single pixel clusters.
For each $\eta$ we associate a corrected value given by the function
\begin{equation}\label{eq:F_eta}
  F(\eta) = \frac{\int^\eta_{-0.5}dN/d\bar{\eta}~d\bar{\eta}}{\int^{0.5}_{-0.5}dN/d\bar{\eta}~d\bar{\eta}} -\frac{1}{2}
\end{equation}
where $\eta$ is in pixel units. The $F(\eta)$ function is shown in Fig.~\ref{fig:eta_function}.
The corrected position is calculated by adding $F(\eta)$ to the 
integer part of the reconstructed position.

\section{Cluster size and position resolution\label{sec:results}}

\begin{figure*}[hbt]
  \begin{center}
    \mbox{
      \subfigure[]{\scalebox{0.33}{
          \epsfig{file=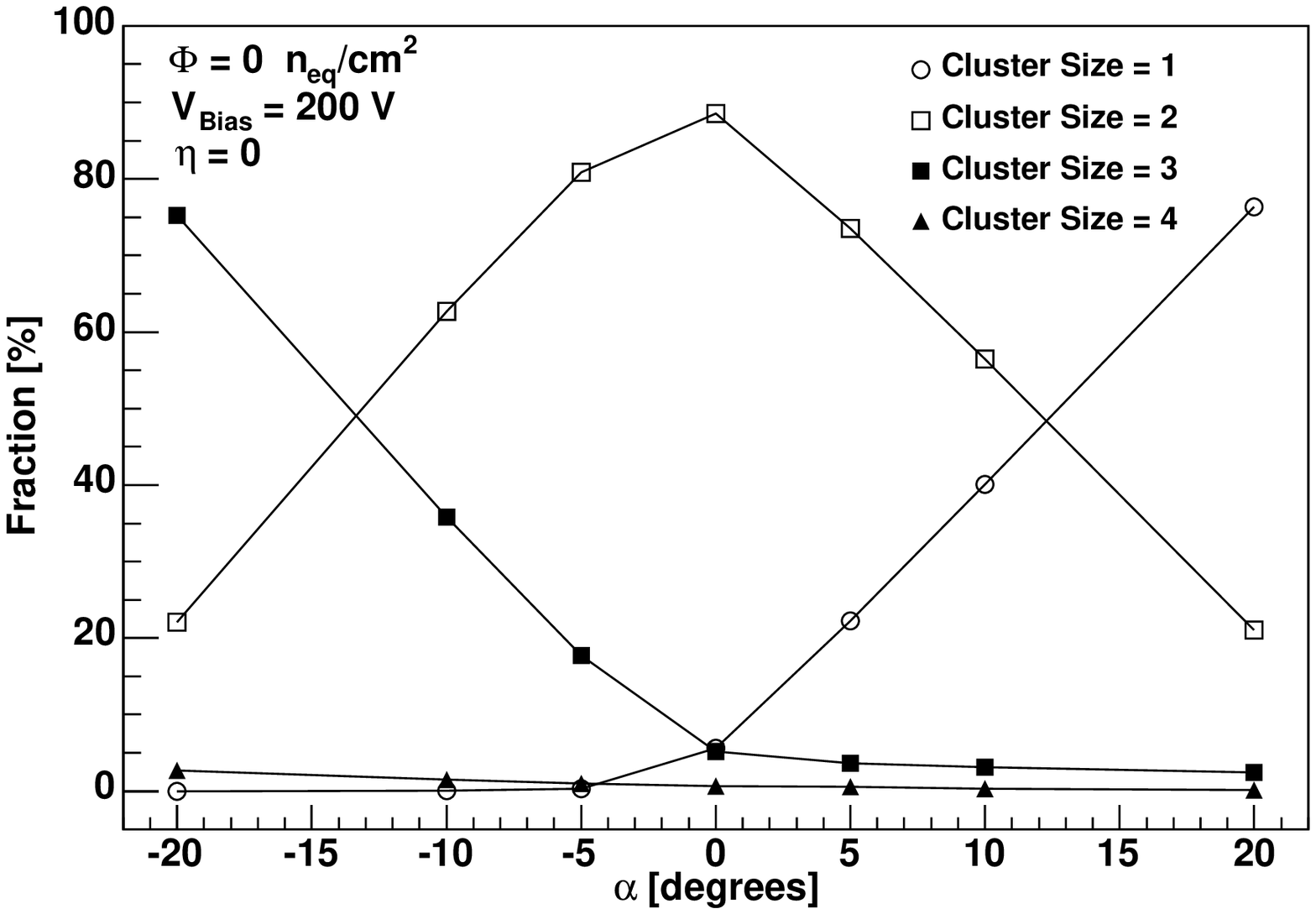,width=\textwidth}
          \label{fig:cl_size_a}
      }}
      \subfigure[]{\scalebox{0.33}{
          \epsfig{file=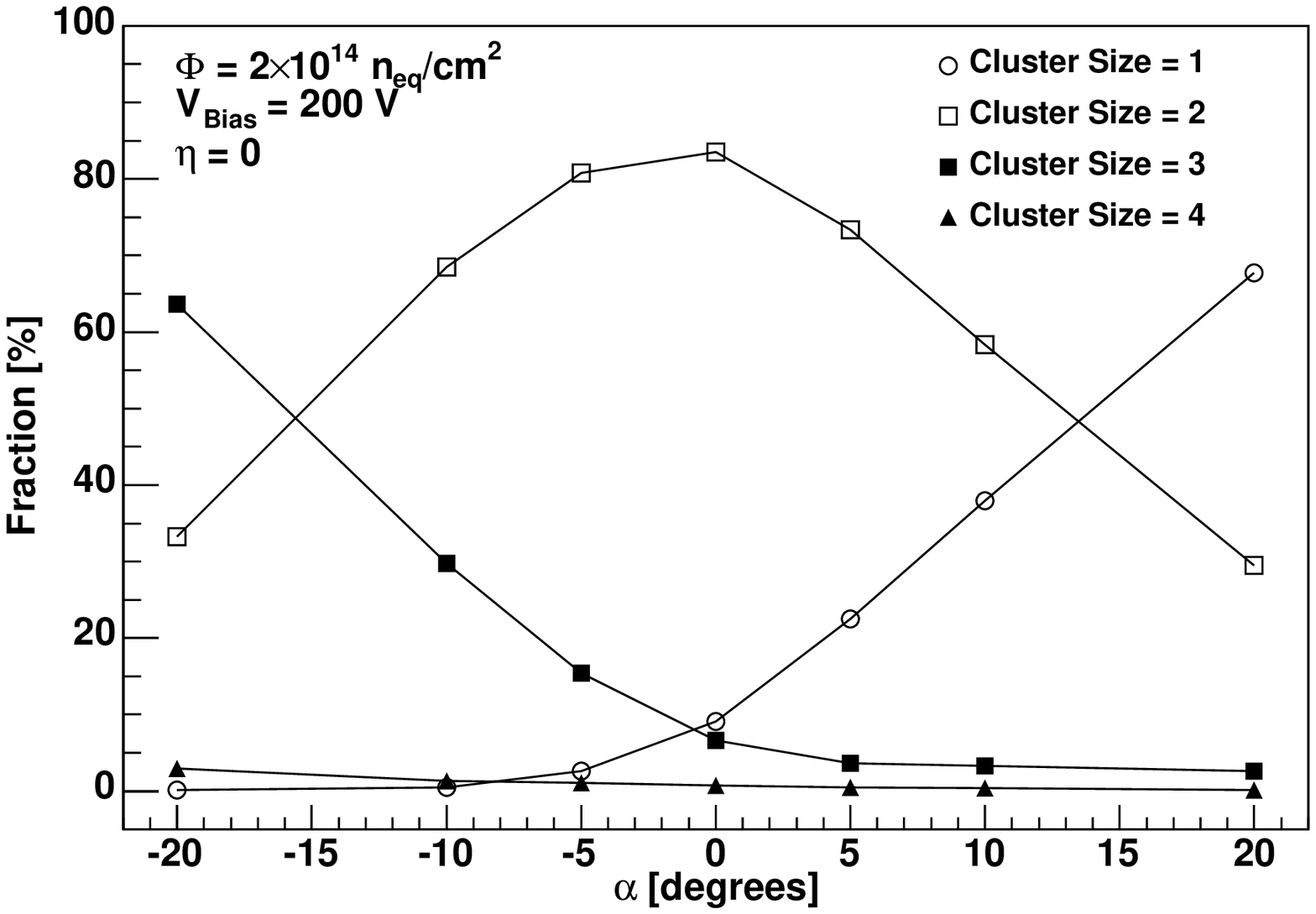,width=\textwidth}
          \label{fig:cl_size_b}
      }}
      \subfigure[]{\scalebox{0.33}{
          \epsfig{file=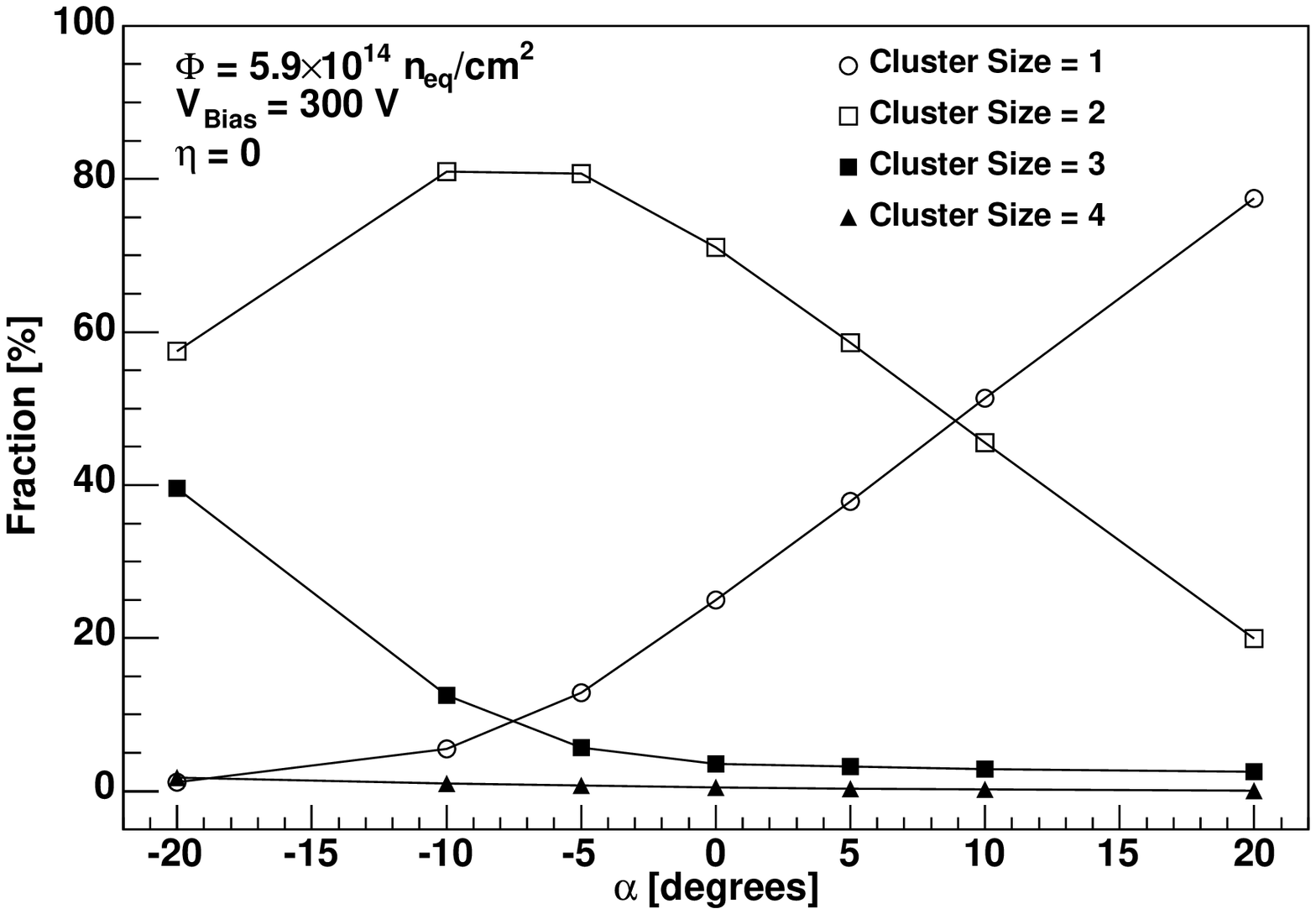,width=\textwidth}
          \label{fig:cl_size_c}
      }}
    }
    \caption{Event fractions for different cluster sizes as function of the impact angle $\alpha$ for 
    an unirradiated sensor (a) and for sensors irradiated to $\Phi = 2 \times 10^{14}$ n$_{\rm{eq}}$/cm$^2$ (b)
    and $\Phi = 5.9 \times 10^{14}$ n$_{\rm{eq}}$/cm$^2$ (c).}
    \label{fig:cluster_sizes}
  \end{center}
\end{figure*}

Figure~\ref{fig:cluster_sizes} shows the fraction of events for different
cluster sizes as function of the impact angle $\alpha$ and for different
irradiation fluences. At negative
angles the Lorentz shift and the geometrical sharing term sum up giving
large $W$ values. For perpendicular tracks ($\alpha=0^\circ$) the total charge width is given 
by the Lorentz shift and then decreases for $\alpha>0^\circ$.
Clusters are larger than one pixel in the majority of cases, however
at high irradiation fluences the higher bias voltage produces a narrower
Lorentz deflection and, consequently, a smaller cluster size.

The combination of carrier trapping and higher bias voltage produces
smaller cluster sizes after heavy irradiation. The average cluster size
for an unirradiated detector with V$_{bias}$=200 V and perpendicular tracks
is 2.0. After a fluence of $\Phi = 5.9 \times 10^{14}$ n$_{\rm{eq}}$/cm$^2$
and V$_{bias}$=300 V the value is 1.8.

The position resolution was estimated by comparing the hit position, $x_{rec}$,
reconstructed as described in Section~\ref{sec:hit_rec} with the true
impact position, $x_{true}$, generated by the simulation. The residuals
were defined as $x_{res} = x_{rec} - x_{true}$ and the position resolution
was given by the width of a Gaussian fit of the residual distribution.
The position resolution for tracks perpendicular to the sensor is summarized in Table~\ref{tab:pos_resolution}
for clusters of two pixels. The position resolution for all events is shown in the last column,
where the position of the two-pixels clusters was $\eta$ corrected. 
The simulation shows that position resolution below 15 $\mu$m can be achieved 
even after heavy irradiation. In addition, the precision can be improved by correcting
the reconstructed position as described in Section~\ref{sec:hit_rec}.
\begin{table*}[hbt]
  \begin{center}
    \begin{tabular}{cc|cc|c}
      \hline
      $\Phi$                 & V$_{bias}$ & Resolution         & Resolution & Total Resolution \\
      (n$_{\rm{eq}}$/cm$^2$) &      (V)   & w/o corr. ($\mu$m) &  ($\mu$m)  & ($\mu$m)       \\
      \hline
      0                 & 200 &  9.3$\pm$0.1 &  9.1$\pm$0.1 &  9.0$\pm$0.1 \\
      2$\times 10^{14}$ & 200 & 13.4$\pm$0.2 & 11.9$\pm$0.2 & 12.1$\pm$0.2 \\
      6$\times 10^{14}$ & 300 & 13.3$\pm$0.1 & 12.3$\pm$0.1 & 12.9$\pm$0.1 \\
      \hline
      \end{tabular}
    \vspace{2mm}
    \caption{Position resolution for $\alpha = 0^\circ$ at different irradiation fluences and
      bias voltages. The third and fourth column show the position resolution of the two-pixel
    clusters without and with $\eta$ correction, respectively. The last column shows the resolution
    for all events, where the position of the two-pixel clusters was $\eta$ corrected.}
    \label{tab:pos_resolution}
  \end{center}
\end{table*}

\begin{figure*}[hbt]
  \begin{center}
    \mbox{
      \subfigure[]{\scalebox{0.45}{
          \epsfig{file=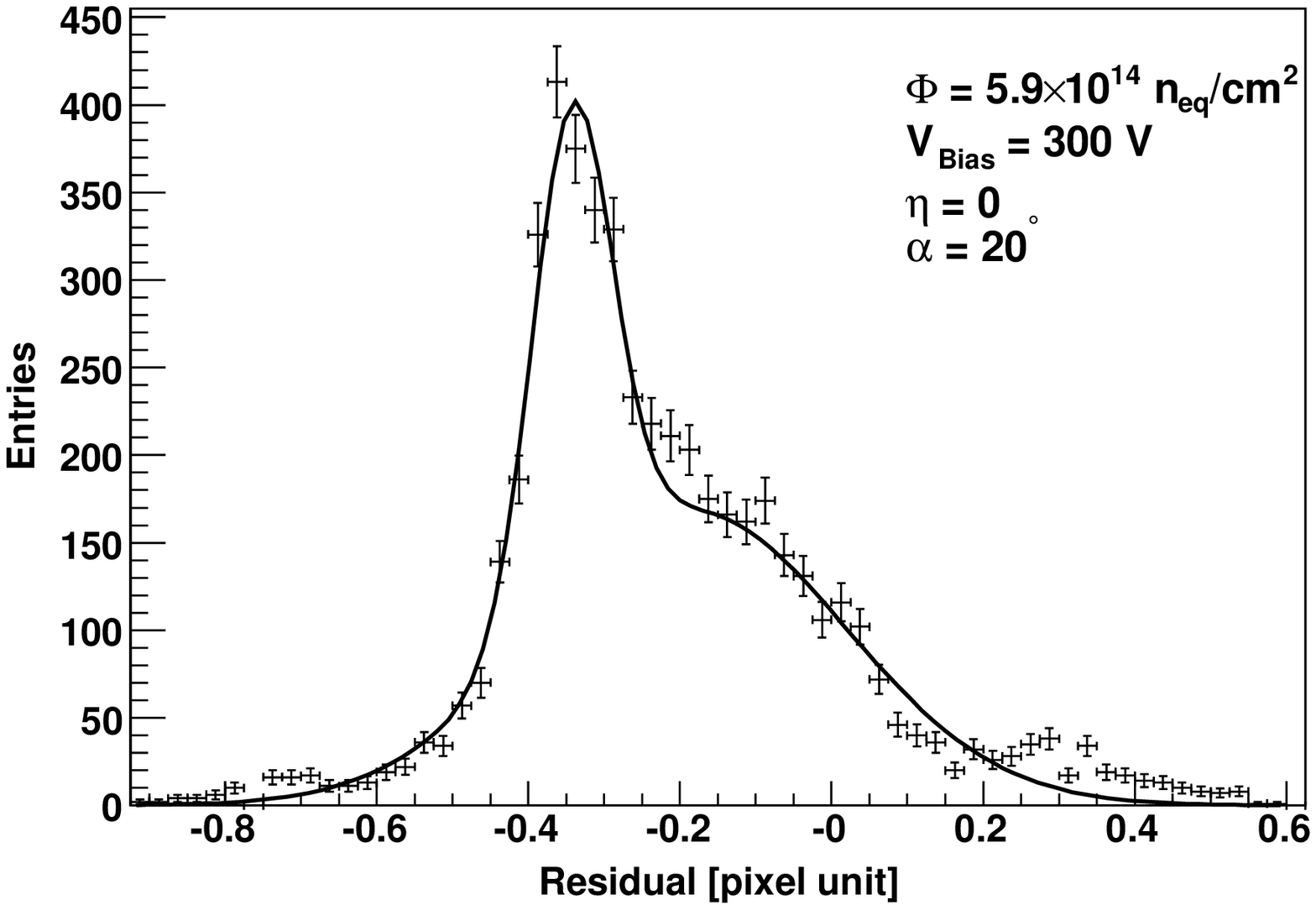,width=\textwidth}
          \label{fig:residuals_a}
      }}
      \subfigure[]{\scalebox{0.45}{
          \epsfig{file=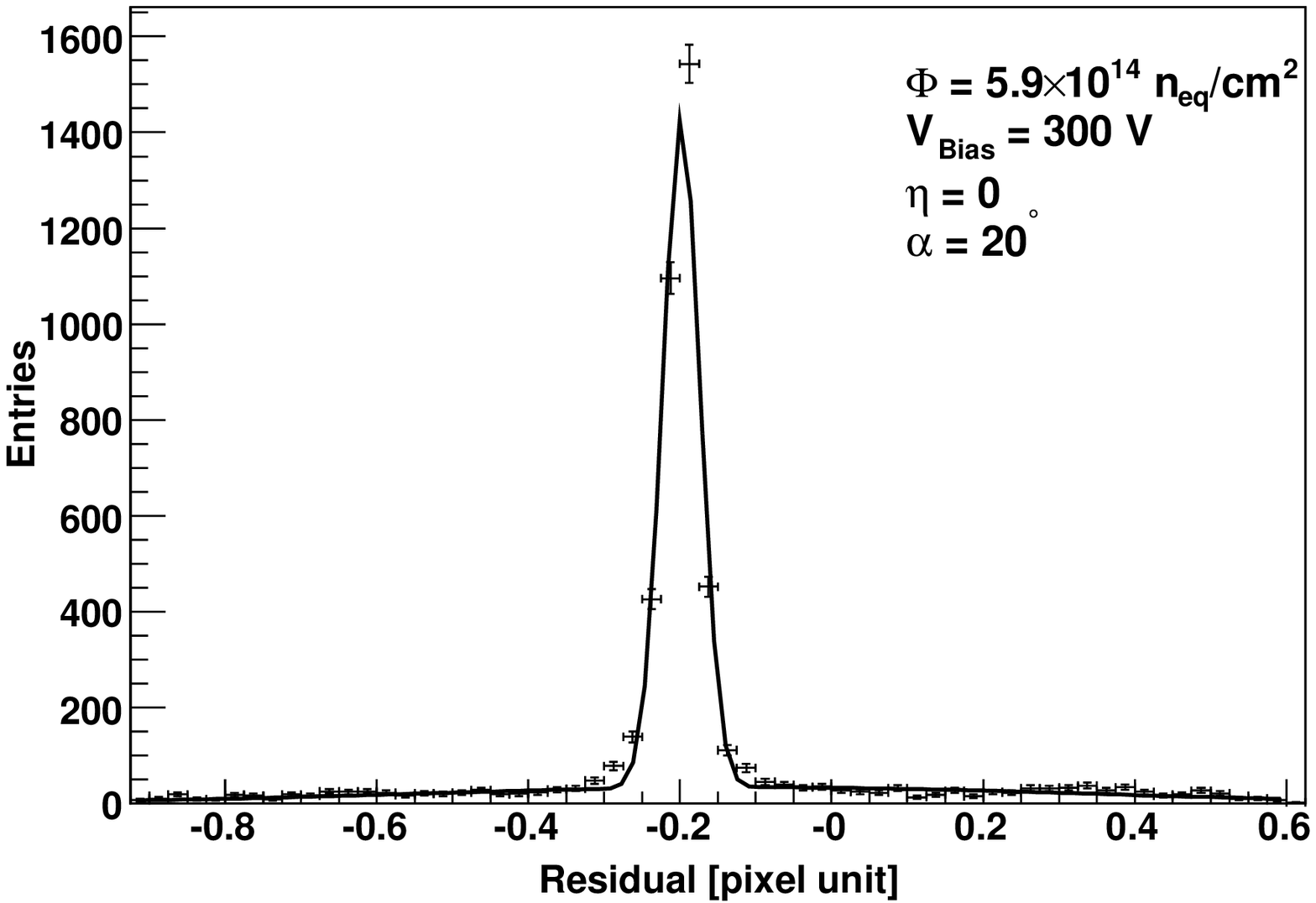,width=\textwidth}
          \label{fig:residuals_b}
	}}
    }
    \caption{Residuals distribution for tracks with $\alpha = 20^\circ$ and a sensor
      irradiated to $\Phi = 5.9 \times 10^{14}$ n$_{\rm{eq}}$/cm$^2$. The
      distributions are calculated without (a) and with $\eta$-corrections (b) 
      and are not corrected for the Lorentz shift due to the magnetic field.
    The simulated data points are represented by the markers and the continuous line is 
    a double-Gaussian fit to the distribution.}
    \label{fig:eta_corr_effect}
  \end{center}
\end{figure*}

The size of the $\eta$ corrections become particularly significant after irradiation 
and for narrow charge widths. Figure~\ref{fig:eta_corr_effect} shows the residuals
distribution for clusters of two pixels, simulated for a sensor irradiated to $\Phi = 5.9 \times 10^{14}$ n$_{\rm{eq}}$/cm$^2$
and for tracks with $\alpha = 20^\circ$. The distribution before correction (Fig.~\ref{fig:residuals_a})
is not described by a single Gaussian and is affected by large systematic errors which depend 
on the interpixel hit position. The distribution in Fig.~\ref{fig:residuals_b} shows that
the systematic errors can be largely reduced by applying the $\eta$-correction.

\section{Conclusions\label{sec:conclusions}}

In this paper a detailed simulation of the silicon pixel sensors 
for the CMS tracker was used to estimate the effects of radiation
damage on the position resolution. The simulation, incorporating a
double junction model of radiation damage and trapping of charge
carriers, provides a good description of the charge collection
measurements in the fluence range from $0.5 \times 10^{14}$ n$_{\rm{eq}}$/cm$^2$
to $5.9 \times 10^{14}$ n$_{\rm{eq}}$/cm$^2$.

The simulation shows that a position resolution below 15 $\mu$m 
can be achieved for perpendicular tracks in the CMS trasverse plane
even after heavy irradiation. In addition, we show that the position
resolution can be improved by applying $\eta$ corrections.



\bibliographystyle{elsart-num}    


\bibliography{refs}             

\end{document}